# Bibliothèques numériques et gamification : panorama et état de l'art

Mathieu Andro (1,2), Imad Saleh (2)

(1) DIST, Institut National de la Recherche Agronomique. mathieu.andro@versailles.inra.fr
(2) Paragraphe, Université Paris 8

## Résumé

Cet article présente un panorama des principaux projets ayant eu recours à la gamification au bénéfice de projets de numérisation et de bibliothèques numériques, que ce soit dans le domaine de l'indexation ou dans celui de la correction de l'OCR. Ce panorama de projets est suivi d'un état de l'art comportant les fonctionnalités, les motivations et la sociologie des contributeurs des projets de gamification et le périmètre de la gamification comparée aux serious games et au crowdsourcing explicite. En conclusion une comparaison entre crowdsourcing explicite et gamification au regard des résultats obtenus est proposée.

<u>Mots clés</u> : numérisation, bibliothèques numériques, tagging, indexation, correction de l'OCR, crowdsourcing, gamification, Games With A purpose, GWAP, Human computation.

## Abstract

**Digital libraries and gamification: overview and state of the art**
This article presents an overview of the main gamification projects for digital libraries, either for tagging or OCR correction. This overview is followed by a state of the art with functionalities, motivations, sociology of contributors and the scope of gamification compared to the serious games and explicit crowdsourcing. Finally a comparison of results between explicit crowdsourcing and gamification is proposed.

## Introduction

Chaque jour, un demi-milliard de personnes à travers le monde jouent à des jeux en ligne pendant au moins une heure  (Eickhoff, 2012). Parmi tous ces jeux, ceux qui ont lieu sur les réseaux sociaux concerneraient pas moins de 120 millions de personnes et le jeu Facebook Farmville attirerait, par exemple, 83 millions de joueurs par mois (Paraschakis, 2013). Les jeux occasionnels (ou "casual games") de type solitaire, puzzle, démineurs, solitaires ou réussites rassembleraient, quant à eux, quelques 200 millions de personnes (Ridge, 2011). Et les Etats-Unis seuls cumuleraient chaque jour, 200 millions d'heures passées sur les jeux vidéos si bien que, dans ces conditions, un américain moyen de 21 ans aurait déjà consacré quelques 10 000 heures de sa vie à des jeux vidéos, soit l'équivalent de 5 années de travail à temps plein sur la base de 40 heures par semaine. (Von Ahn, 2008).

Le temps croissant que les internautes consacrent à ces jeux pourrait être réutilisé à des fins productives, en particulier en leur proposant de réaliser, sous la forme de jeux en ligne, des micro-tâches qui demeurent impossibles à effectuer par des ordinateurs alors qu'elles le sont pour le cerveau humain. C'est tout l'enjeu de la gamification, une forme de crowdsourcing qui externalise le travail auprès de foules d'internautes, mais dont la spécificité est de le faire sous la forme de jeux. Le potentiel de la gamification suscite l'enthousiasme de bien des auteurs. Luis Von Ahn, l'un des inventeurs du célèbre reCAPTCHA, modèle de crowdsourcing implicite, prétendait ainsi que l'intégralité des images de Google Images (425 millions d'images) aurait pu être indexée grâce au jeu ESP Game en seulement 31 jours par 5000 personnes qui auraient joué continuellement à son jeu emblématique de la gamification (Von Ahn, 2008). Cet enthousiasme pourrait également trouver un écho économique si l'on considère les projections rapportées dans une récente étude (Ollikainen, 2013).

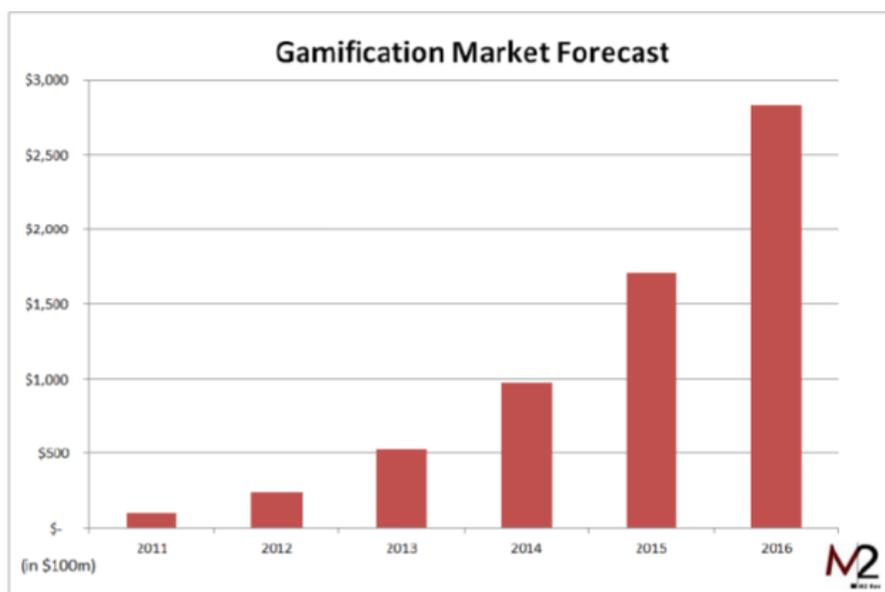

**Evolution du marché de la gamification (d'après Ollikainen, 2013)**

A l'instar du crowdsourcing dont la gamification est l'une des formes, l'abolition de la séparation entre travail et loisir, entre production et consommation qui est sous-jacente dans ce modèle économique revêt un caractère conceptuellement quasi révolutionnaire susceptible de faire echo à des idéologies aussi diamétralement opposées que le sont le marxisme avec son émulation socialiste et son stakhanivisme et le libéralisme avec ses principes de libertés individuelles, de "main invisible" et de "fun at work" (Andro, 2014). Le travail lui-même sous sa forme actuelle pourrait d'ailleurs, à plus d'un titre, être considéré, à l'instar des jeux, comme une succession de challenges, avec des épreuves, des quêtes, des changements de niveaux, des points et des bonus et, par conséquent, être stimulé par les mêmes ressorts et mécaniques que ceux utilisés dans un tout autre cadre, celui du jeu vidéo.

La notion de "gamification" qui correspond à ce modèle économique et qui pourrait être traduite en français par "ludification", a été proposée, dès 2002, par Nick Pelling. Elle pourrait être définie comme le fait d'appliquer des éléments de design et de psychologie et des mécanismes du jeu video dans d'autres contextes (Deterding, 2011 b). Le terme de "Human Computation" proposé, 3 ans plus tard, par Luis Von Ahn et qui lui est très proche, pourrait être défini comme la mobilisation de l'intelligence humaine collective, par des jeux, afin de résoudre des problèmes que les ordinateurs ne peuvent pas encore résoudre ou qui ne peuvent l'être par des groupes trop limités d'humains (Quinn, 2011). Les cerveaux humains pourraient ainsi être considérés comme des processeurs au sein d'un système distribué au sein duquel chacun contribuerait à générer un calcul massif (Von Ahn, 2006). On parle ainsi de "Human Computation" pour désigner ce mouvement visant à remplacer les algorithmes informatiques par l'intelligence humaine. La notion de "Games with a purpose" ou "GWAP" désigne, quant à elle, les jeux avec une finalité utile et productive qui résultent de ce mouvement.

Des applications ont déjà été trouvées dans la sécurité, la vidéosurveillance, la traduction, les résumés de textes, l'éducation ou l'indexation des images. Dans un contexte de recherche de plaisir et de ludification de la culture, cette démarche innovante pourrait également trouver des applications dans le domaine des bibliothèques numériques et de la numérisation du patrimoine culturel. Ce sera l'objet de cet article.

# 1- Panorama de quelques projets de gamification représentatifs dans le domaine des bibliothèques numériques

## 1.1- L'indexation en jouant

Bien avant l'apparition des premières applications de la gamification dans le domaine des bibliothèques numériques, certaines expérimentations, relatives à l'indexation d'images en particulier, méritent ici d'être brièvement rapportées car elles les ont fortement inspirées.

Ainsi, entre 2006 et 2011, Google Image Labeler fut l'une des premières et des plus célèbres expérimentations dans le domaine de la gamification. Développé sous la direction de Luis Von Ahn, Google Image Labeler consistait simplement à proposer la même image à deux internautes dans le monde qui marquaient des points s'ils parvenaient à saisir le même mot clé pour l'indexer. Grâce à cette confrontation des saisies collectées, Google est ainsi parvenu à ajouter, en 2008, plus de 50 millions de mots clés aux images indexées par Google Image. Cette expérimentation a ensuite été prolongée par ESP Game rebaptisé par la suite GWAP et qui ajoutait des mots tabous interdits à l'indexation, issus de parties précédentes afin de forcer les internautes à produire une indexation plus spécifique. Le jeu Peekaboom qui lui succéda avait, quant à lui, pour finalité de localiser précisément à quelle partie de l'image se rapportait tel ou tel mot clé collecté via ESP Game, un premier joueur ayant pour mission de le faire deviner à un second en dévoilant, avec ses clics, les parties correspondantes de l'image. Par la suite, le jeu KissKissban (KKB) a repris le concept de ESP Game en y ajoutant un 3ème joueur dont le but était d'empêcher le binôme de joueurs de marquer des points en proposant des mots clés bloquants inconnus d'eux.

Art Collector est un jeu Facebook développé par le Swedish Open Cultural Heritage (SOCH) à partir de près de 100 000 images collectées sur plusieurs sites du patrimoine culturel suédois. Il fait à la fois appel à l'esprit de compétition (résultats, challenges, tableau des meilleurs joueurs), mais aussi à la collaboration (partager un trophée) et à la communication (recherche d'amis, notifications) (Paraschakis, 2014). Dans une première partie du jeu, chaque joueur cherche à constituer la plus grande collection d'images. Pour s'approprier une image parmi les quatre qui lui sont proposées, il suffit d'être celui qui a proposé le plus de tags pour la décrire. Ensuite, dans une deuxième partie, chaque joueur cherche à s'approprier des images au sein des collections des autres collectionneurs en devinant plus de la moitié de leurs tags. Ainsi, grâce au premier round, les institutions culturelles obtiennent des indexations en nombre et grâce au second, elles font valider ces indexations en les confrontant entre elles.

Metadata Games est une expérimentation commencée avec la Bibliothèque du Dartmouth College puis l'Université de Washington, la Bibliothèque Publique de Boston, l'université de Buffalo, l'UC-Santa Cruz et Hong Kong. Les jeux open source, développés en HTML5, sur le site proposent de taguer, de renseigner des champs, de proposer de trouver le même tag qu'un autre internaute sur le modèle de Google Image Labeler ou encore de faire deviner à un autre internaute avec l'aide de tags une bonne image parmi 12 images. (Flanagan, 2012).

museumgam.es est un projet de jeux développés sous WordPress avec des développements PHP utilisent les données du Science Museum et du Powerhouse Museum via leurs API et proposent aux internautes de taguer des images, d'ajouter des informations aux objets, ou d'aider un jeune conservateur à réindexer toutes les données perdues accidentellement.

Nous pourrions également évoquer SaveMyHeritage, un jeu sur les réseaux sociaux mettant en compétition des internautes pour taguer des photographies. (Havenga, 2012) ou encore Alum Tag, développé par la Rauner Special Collections Library du Dartmouth College et consistant à indexer des photographies, ou enfin, Tag! You're it!, développé par le Brooklyn Museum pour l'indexation des objets.

## 1.2- La correction de l'OCR en jouant

A l'issue de la numérisation des textes imprimés par les bibliothèques, un traitement informatique de reconnaissance optique de caractères (OCR) va identifier à quel caractère correspond la photographie de tel caractère et parfois générer des erreurs de reconnaissance qui peuvent ensuite être corrigées par des sociétés ou par des foules d'internautes grâce au crowdsourcing explicite ou encore en ayant recours à des jeux.

Développé par la société Microtask depuis 2011, Digitalkoot, dont le nom est inspiré du "talkoot", un mode traditionnel et collectif finlandais de construction des maisons fondé sur l'entraide, propose aux internautes de corriger de l'OCR brute issue de campagnes de numérisation de la Bibliothèque Nationale de Finlande sous la forme de jeux. Le premier jeu ("Mole Hunt" ou chasse aux taupes) consiste à exposer à l'internaute 2 mots différents, l'un correspond à l'image numérisée du mot et le second à la suggestion du logiciel d'OCR. L'internaute doit alors déterminer le plus rapidement possible si ce sont bien les mêmes mots et valider ainsi les résultats de l'OCR brut.

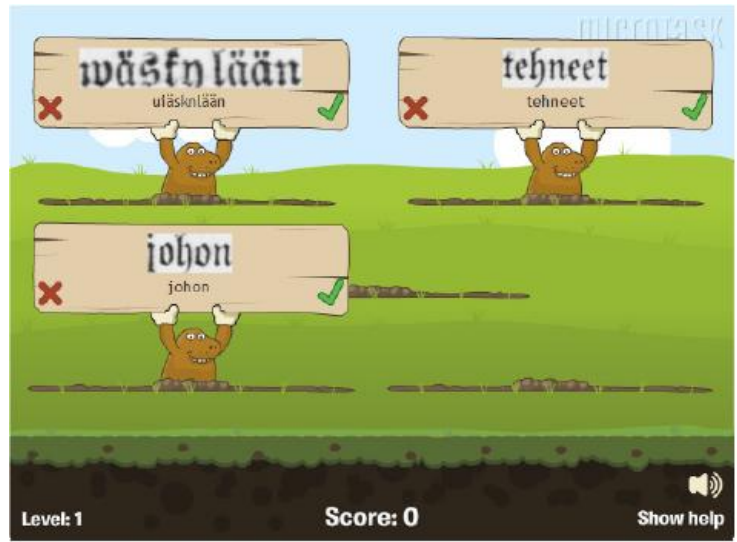
**Figure 1. Capture d'écran du jeu Mole Hunt**

Le deuxième jeu ("Mole Bridge" ou pont de taupes) consiste à transcrire les images numérisées de mots. A chaque bonne réponse, une brique du pont à construire, pour aider les taupes à traverser une rivière, s'ajoute aux précédentes. Par contre, si le mot saisi ne correspond pas, une brique du pont explosera. Il faut construire le pont en évitant au maximum que les taupes ne tombent dans la rivière. A l'instar de n'importe quel jeu d'arcades, les décors, les vitesses, les distances… changent en fonction des changement de niveaux, ce qui stimule les joueurs à poursuivre leur progression et à continuer de jouer.

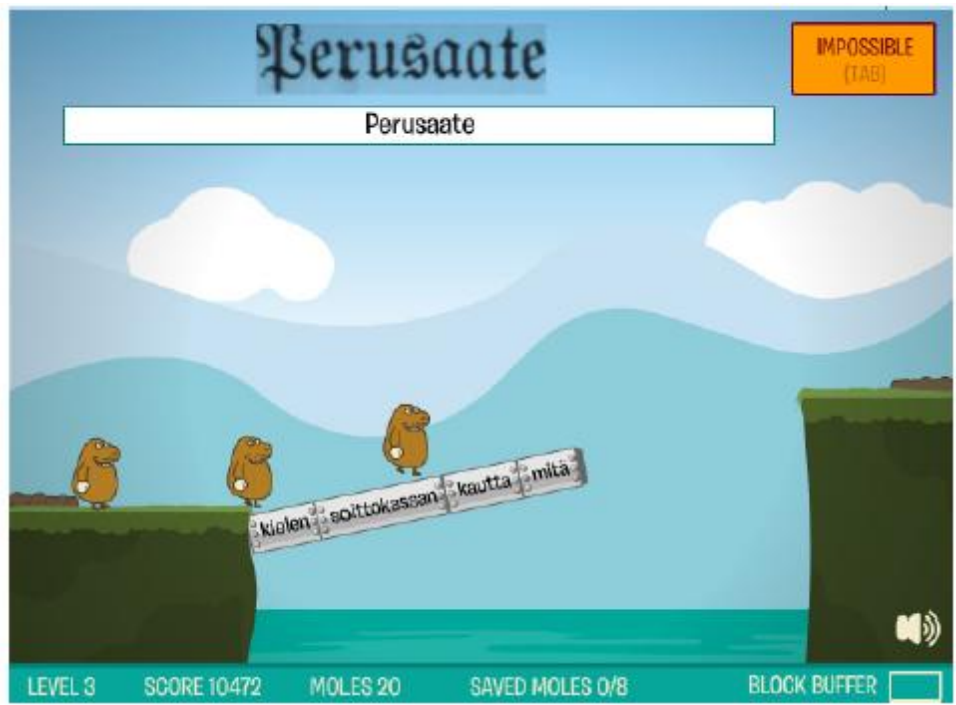
**Figure 2. Capture d'écran du jeu Mole Bridge**

Les compétences, les aptitudes et la bonne volonté des nouveaux contributeurs sont préalablement vérifiées en leur soumettant des mots tests et en les confinant dans un "bac à sable" afin de vérifier qu'ils ne sont pas de malveillants. La qualité des corrections est également garantie par la méthode classique de confrontation les saisies de plusieurs internautes. Au final, la qualité de l'OCR corrigé obtenue avec le jeu serait de 99 % (sur 2 articles de 1467 et 516 mots, 14 erreurs et 1 erreur ont respectivement été trouvées) alors que l'OCR brute de départ aurait été une qualité moyenne de 85 %.

En février 2012, c'est à dire un an après son ouverture le 8 février 2011, le site comptabilisait déjà 5473 heures de jeu, soit une moyenne de près de 15 heures de jeu par jour. En octobre 2012, il avait rassemblé un total de 109 321 contributeurs ayant réalisé 8 024 530 microtâches, soit une moyenne de 13 717 microtâches par jour. En considérant qu'il y a entre 220 et 260 mots par page de livre numérisé, on peut ainsi estimer qu'en octobre 2012, entre 30 000 et 37 000 pages de livres avaient ainsi été corrigées grâce à ce jeu, soit l'équivalent de 154 à 182 livres de 250 pages. En considérant, à présent, que la correction d'une page de texte océrisée par un prestataire varie entre 1 € et 1,5 € la page, le projet aurait ainsi évité de dépenser, en octobre 2012, entre 31 000 et 55 000 € de budget en correction de l'OCR, soit en moyenne, 1476 € à 2619 € par mois. Bien que nous ignorions quel a été le coût précis de développement du projet, il est probable que ce coût soit donc amorti prochainement.

Mais le résultat le plus notable du projet est qu'une partie non négligeable de la population Finlandaise qui compte un peu plus de 5 millions d'habitants a participé à ce projet. Si on rapporte les 109 321 contributeurs déjà évoqués, qui ne sont cependant pas tous finlandais, à ces 5 millions d'habitants, c'est près d'un finlandais sur 46 qui aurait participé. La communication autour du projet fait d'ailleurs appel au sentiment patriotique finlandais des internautes en les invitant à "sauvegarder la culture finlandaise". Une campagne de communication, dans la presse (Wired, New York Times), les blogs et même à la télévision, ont permis de faire largement connaître ces jeux astucieux. Le fait de permettre une authentification directement via Facebook, utilisé par 98 % des internautes sur le site, a également amené a faire largement connaître le site sur les réseaux sociaux. Ainsi, pendant la première semaine d'ouverture, 1756 personnes sont venues sur le site par ce moyen. Un tiers des internautes, ont amenés leurs amis sur Facebook à visiter le site pendant la première semaine. 99 % des visiteurs seraient ainsi venus de Facebook et seulement 1 % via Google.

Dans le domaine de la correction de l'OCR avec l'aide de la gamification, nous pouvons également évoquer le projet COoperative eNgine for Correction of ExtRacted Text (CONCERT) développé en 2009, dans le cadre du projet européen IMPACT par IBM Israël en partenariat avec des institutions et des sociétés comme ABBY et qui a développé des jeux sur les réseaux sociaux et sur smartphones, ou encore le jeu Facebook TypeAttack, résultat de la collaboration entre l'Université de et la Bibliothèque de Singapour et qui autorise une compétition entre amis du même réseau social sur Facebook, ou enfin le jeu Word Soup Game qui propose aux internautes d'identifier des mots, issus de l'OCR brute, au sein d'une grille de lettres désordonnées appelée soupe de mots.

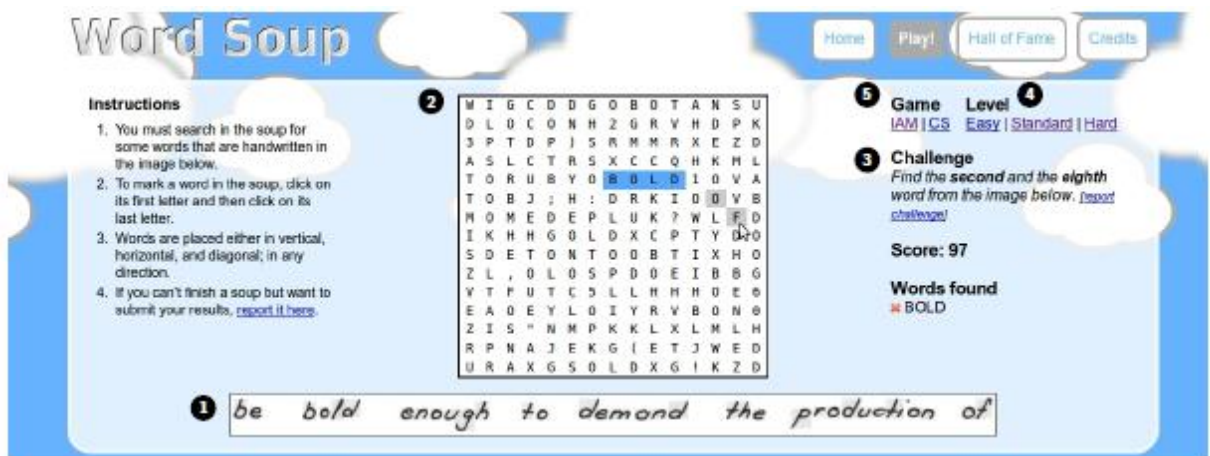

**Capture d'écran de Word Soup d'après (Alabau, 2012)**

# 2- Etat de l'art de la gamification en bibliothèques numériques

## 2.1- Fonctionnalités spécifiques de la gamification

A partir du panorama qui précède et de la littérature (Hamari, 2014), les fonctionnalités suivantes caractéristiques de la gamification peuvent être identifiées :

- fonctionnalités sociales : la possibilité de partager et de faire connaître le jeu via son réseau social de manière virale, d'inviter des internautes à jouer directement depuis son réseau social, puis d'échanger avec eux à travers le jeu.
- crowdfunding : les micro paiements permettant aux internautes d'encourager financièrement le développement des jeux.
- stimulation de la motivation : statistiques, nombre de points, compétition, challenges et défis, classements des joueurs, médailles, niveaux, limitation du temps, présence de mots tabous pour contraindre les internautes à proposer des mots clés plus spécifiques.

L'affichage du top des meilleurs joueurs sur le mois, sur la semaine, ou sur la journée permettrait d'encourager la participation des joueurs qui, sur une durée plus limitée, peuvent espérer monter sur le podium (Von Ahn, 2008). En complément, il serait également possible d'afficher le top des meilleurs joueurs en fonction d'une zone géographique afin d'encourager les joueurs à chercher à améliorer leur classement sur leur ville, leur région, leur pays, ou leur continent et de mieux adapter les objectifs aux types de joueurs (Ridge, 2011).

## 2.2- Motivations et sociologie des "joueurs-travailleurs"

A la différence du crowdsourcing explicite classique, le recours à la gamification semble faire appel d'avantage à des motivations extrinsèques de type reconnaissance sur les réseaux sociaux et récompenses virtuelles ou réelles, et à des motivations intrinsèques liées à la distraction et au plaisir de jouer, plutôt qu'à des besoin de développement personnel et d'acquisition de compétences comme ceux que l'on peut acquérir en transcrivant des manuscrits, par exemple.

La sociologie des contributeurs serait également sensiblement différente. Contrairement à la correction participative de l'OCR sous sa forme plus classique de crowdsourcing explicite et qui, comme c'est par exemple le cas pour le projet australien TROVE, toucherait plutôt des retraités passionnés de généalogie et d'histoire locale, l'approche gamification permettrait de toucher des populations plus jeunes. Ainsi, concernant le jeu Art Collector, la classe de joueurs la plus nombreuse serait celle des 25-34 ans et concernant Digitalkoot, les volontaires seraient majoritairement âgés de 25 à 44 ans.

On note également parfois une surreprésentation des hommes par rapport aux femmes (10 % de plus pour Art Collector). Néanmoins, les femmes domineraient sur museumgam.es. Et s'agissant de Digitalkoot, si la moitié des utilisateurs seraient des hommes, les femmes passeraient en moyenne plus de temps sur le site qu'eux (13 minutes pour les femmes, un peu plus de 6 minutes pour les hommes) et elles réaliseraient 54 % des tâches (Chrons, 2011). Néanmoins, le top 4 des plus gros contributeurs seraient des hommes, le plus important d'entre eux ayant fourni 101 heures de travail pour réaliser 75 000 microtâches à lui tout seul. Ainsi, seulement 1 % des internautes réalisent plus du tiers du travail.

Les joueurs masculins pourraient avoir d'avantage tendance à évaluer leur performances que les femmes qui elles seraient plutôt attirées par le caractère relationnel des jeux (McCarthy, 2012). Elles seraient d'avantage attirées par les jeux de puzzle, d'aventure, de combat et de gestion, tandis que les hommes préféreraient les jeux de tir, de sport, de rôle ou de stratégie. Une enquête de 2006 de la société PopCap, révélerait ainsi que 76 % des joueurs à des jeux occasionnels (casual games) de type solitaire, puzzle, démineurs, solitaires, réussites… seraient des femmes dont la moyenne d'âge serait de 48 ans (Ridge, 2011).

## 2.3- Le périmètre de la gamification

Le fait d'attribuer des points pour tout et n'importe quoi sur le web ne doit pas être confondu avec de la gamification. On devrait, à ce sujet, plutôt parler de "pointification". Les jeux, quant à eux, demandent un espace et un temps définis, ils obéissent à des règles, à des objectifs, sont volontaires et satisfont des motivations plutôt intrinsèques.

### 2.3.1- Serious games et gamification

La gamification peut être considérée comme étant au croisement entre crowdsourcing et serious games (Harris, 2013). Néanmoins, si, à l'instar des serious games, la gamification est une forme de jeux avec une finalité productive (Games With A Prupose), elle s'en distingue dans la mesure où cette finalité ne vise pas un développement personnel mais la réalisation d'objectifs extérieurs à soi (indexer une collection numérisée, corriger l'OCR de textes numérisés). Par ailleurs, la finalité des serious game est utilitaire et "sérieuse" pour l'usager alors qu'avec la gamification, il cherche principalement à se distraire, même si les données qu'il produit en se distrayant ont une autre finalité et un usage qui peuvent être tout aussi sérieux pour la bibliothèque qui l'a développé. Enfin, contrairement aux serious game, la gamification fait appel à des microtâches et ne propose pas un scénario très linéaire et complet.

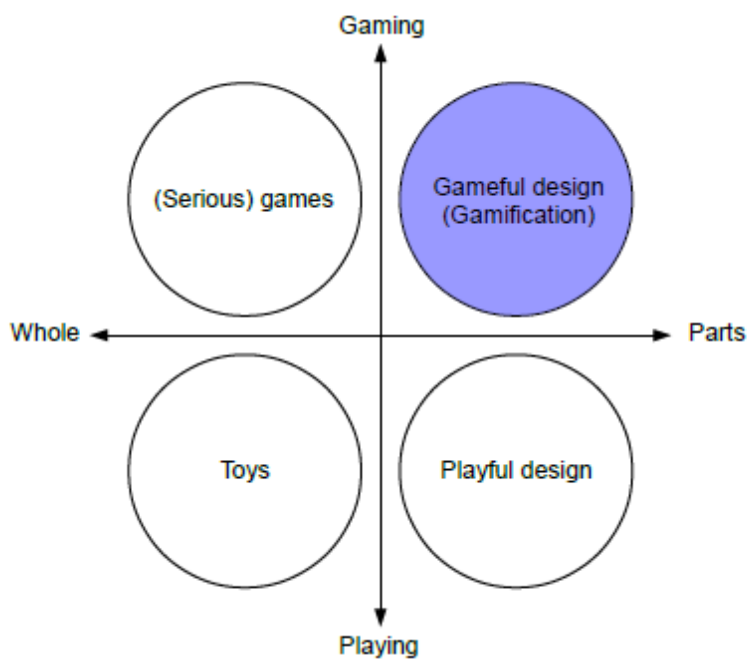

**La différence entre gamification et serious games d'après (Groh, 2012) reprenant (Deterding, 2011 a)**

### 2.3.2- Crowdsourcing explicite et gamification

On distingue le crowdsourcing explicite qui fait appel, de manière désormais classique, à des internautes volontaires qui peuvent être bénévoles (Wikisource, TROVE) ou rémunérés (Amazon Mechanical Turk Marketplace), du crowdsourcing implicite qui fait appel au travail involontaire et généralement inconscient des internautes (reCAPTCHA). La gamification fait appel, quant à elle, à des internautes qui cherchent à jouer tout en aidant un projet ou qui cherchent à aider un projet tout en jouant. Elle est donc à mi chemin entre crowdsourcing explicite et implicite, à mi chemin entre contribution volontaire et altruiste et contribution involontaire et ludique.

Concernant la correction de l'OCR, par exemple, le crowdsourcing explicite permet une correction des textes dans le contexte, ce qui permet un développement personnel des internautes qui peuvent ainsi en profiter pour lire des textes qui les intéressent. Ce type de développement personnel n'existe guère avec la gamification. Par ailleurs, le crowdsourcing explicite serait beaucoup moins coûteux et moins longs à mettre en place que les jeux, il y serait plus facile de maintenir la participation des contributeurs et bénéficierait d'une meilleure image éthique (Sabou, 2013).

Par contre, la gamification permettrait d'obtenir de meilleurs résultats en termes de participation (Chrons, 2011), aurait un coût à la tâche très légèrement inférieur, aurait une qualité de production plus élevée, favoriserait moins la fraude (Sabou, 2013), serait plus rapide, permettrait de faire faire des tâches plus complexes, et mobiliserait une variété plus diversifiée de profils (Harris, 2013). Par ailleurs, les jeux permettraient d'obtenir d'avantage de mots clés par personne. Ainsi, entre une correction de l'OCR classique et une correction via Digitalkoot, on obtiendrait 20 % de participation en plus (McCarthy, 2012). Ainsi, on obtiendrait en moyenne 6 tags par visiteur avec le projet Flickr de la Bibliothèque du Congrès contre une moyenne de 84 tags par visiteur avec le jeu Tiltfactor Metadata Game (Flanagan, 2012). En effet, la gamification tient compte du fait que seule une infime minorité d'internautes participe à Wikipedia ou à d'autres projets de crowdsourcing explicite philanthropique et qu'il est, par conséquent, plus opportun et astucieux, de recycler l'énergie des internautes dans leurs activités de jeux sur le web. C'est d'autant plus vrai que les individus ont parfois des difficultés à passer une partie importante de leur temps sur des tâches ingrates alors qu'ils ont, au contraire, parfois des difficultés à cesser de jouer sur ordinateur. (Chrons, 2011).

## Conclusions

Nous ne sommes pas parvenus à identifier de projets faisant appel à la gamification dans les bibliothèques de France. De manière générale, le développement du crowdsourcing y demeure d'ailleurs encore relativement peu développé, en dehors de quelques initiatives avec Wikisource, du projet Ozalid mené à la Bibliothèque nationale de France ou des projets de numérisation à la demande par crowdfunding comme Ebooks on Demand et Numalire. Le difficile développement du crowdsourcing dans les bibliothèques de France s'explique probablement par des difficultés idéologiques et culturelles à externaliser auprès de foules d'amateurs privés d'activités actuellement prises en charge par des professionnels du secteur public.

Reprenant les calculs effectués pour Digitalkoot et les estimations (Zarndt, 2014), nous comparons les coûts non dépensés en correction d'OCR entre deux projets de correction participative de l'OCR ayant une envergure comparable : le projet de crowdsourcing explicite California Digital Newspaper Collection et le projet de gamification Digitalkoot :

| Projet | Type de crowdsourcing | Coût non dépensé |
|---|---|---|
| California Digital Newspaper Collection (fin 2011-) | Crowdsourcing explicite | 53 130 $ cumulés en juin 2014 soit de 1 432 € à 1 629 € par mois |
| Digitalkoot (février 2011-) | Gamification | Entre 31 000 et 55 000 € cumulés en octobre 2012 soit de 1476 € à 2619 € par mois par |

| | | mois |
|---|---|---|

**Tableau 1. Estimation du coût non dépensé en prestations de correction de l'OCR par CDND et Digitalkoot**

Afin de comparer les résultats obtenus avec chacun de ces deux types de projets, il aurait été nécessaire de les rapporter à leurs coûts de développement, d'administration, de maintenance, d'hébergement, de communication, de community management, de réintégration des données produites… En l'état, il reste donc encore difficile de juger de la compétitivité de la gamification par rapport aux formes plus classiques de crowdsourcing. Cela pourra faire l'objet de recherche ultérieure si nous parvenons à obtenir les données manquantes.

Le développement de tels jeux peut avoir un coût non négligeable pour les bibliothèques. Dans ces conditions, la mutualisation de jeux pour l'indexation ou la correction de l'OCR au bénéfice de plusieurs bibliothèques numériques serait nécessaire. Elle pourrait se faire dans le cadre d'un projet européen, par exemple. Elle permettrait de partager les coûts et de rassembler des moyens supérieurs afin de réaliser des jeux de bien meilleure qualité dans le cadre de projets plus pérennes et susceptibles de capter un nombre plus important de joueurs contributeurs. Avec l'abandon annoncé de reCAPTCHA sous sa forme actuelle par Google, la mutualisation d'un Captcha éthique au profit des bibliothèques numériques pourrait également être une solution intéressante à proposer.

En attendant, le "marché" encore disponible pour le recours explicite au bénévolat pourrait commencer à se resserrer à cause de la multiplication des projets et rien n'indique que les générations futures de retraités qui forment une part très importante des contributeurs de projets de crowdsourcing explicite comme TROVE continueront dans l'avenir à se passionner pour la généalogie et l'histoire locale et continueront à corriger l'OCR. Dans ces conditions, miser sur la gamification plutôt que sur le crowdsourcing explicite pourrait s'avérer relativement judicieux.

# Bibliographie